\documentclass[preprint,10pt]{aastex}

\usepackage{relsize}
\usepackage{upgreek}
\usepackage{amsmath}
\usepackage{natbib}
\usepackage{url}

\usepackage{graphicx}
\usepackage{txfonts}
\def\1E0102{1E\,0102}

\newcommand\ha{H$\alpha$}
\newcommand{\kms}{\ensuremath{\mathrm{km}\,\mathrm{s}^{-1}}}
\newcommand\clii{[Cl\,\textsc{ii}]}
\newcommand\sii{[S\,\textsc{ii}]}
\newcommand\siii{[S\,\textsc{iii}]}
\newcommand\oi{[O\,\textsc{i}]}
\newcommand\oii{[O\,\textsc{ii}]}
\newcommand\oiii{[O\,\textsc{iii}]}
\newcommand\nii{[N\,\textsc{ii}]}
\newcommand\neiii{[Ne\,\textsc{iii}]}
\newcommand\ariii{[Ar\,\textsc{iii}]}
\newcommand\caii{[Ca\,\textsc{ii}]}

\begin{document} 

   \title{Integral Field Spectroscopy of Supernova Remnant 1E0102-7219 Reveals Fast-moving Hydrogen and Sulfur-rich Ejecta}

\author{Ivo R.~Seitenzahl\altaffilmark{1,2,3}, Fr{\'e}d{\'e}ric P.~A. Vogt\altaffilmark{4\dagger}, Jason P. Terry\altaffilmark{5}, Parviz Ghavamian\altaffilmark{6}, Michael A. Dopita\altaffilmark{2}, Ashley J. Ruiter\altaffilmark{1,2,3}, Tuguldur Sukhbold\altaffilmark{7,8}
}

\altaffiltext{1}{School of Physical, Environmental and Mathematical Sciences, University of New South Wales, Australian Defence Force Academy, Canberra, ACT 2600, Australia. i.seitenzahl@adfa.edu.au}
\altaffiltext{2}{Research School of Astronomy and Astrophysics, Australian National University, Canberra, ACT 2611, Australia.}
\altaffiltext{3}{ARC Centre for All-sky Astrophysics (CAASTRO).}
\altaffiltext{4}{European Southern Observatory, Av. Alonso de C\'ordova 3107, 763 0355 Vitacura, Santiago, Chile.}
\altaffiltext{$\dagger$}{ESO Fellow.}
\altaffiltext{5}{Department of Physics and Astronomy, University of Georgia, USA.}
\altaffiltext{6}{Department of Physics, Astronomy and Geosciences,
  Towson University, Towson, MD, 21252, USA.}
\altaffiltext{7}{Department of Astronomy, The Ohio State University, Columbus, OH, 43210, USA.}
\altaffiltext{8}{Center for Cosmology and AstroParticle Physics, The Ohio State University, Columbus, OH, 43210, USA.}

 
\begin{abstract}
We study the optical emission from heavy element ejecta in the oxygen-rich young supernova remnant (SNR) \1E0102.2-7219 (\1E0102) in the Small Magellanic Cloud.
We have used the Multi-Unit Spectroscopic Explorer (MUSE) optical integral field spectrograph at the Very Large Telescope (VLT) on Cerro Paranal and the wide field spectrograph (WiFeS) at the ANU 2.3\,m telescope at Siding Spring Observatory to obtain deep observations of 1E\,0102. Our observations cover the entire extent of the remnant from below 3500\AA\ to 9350\AA.
Our observations unambiguously reveal the presence of fast-moving ejecta emitting in \sii, \siii, \ariii, and \clii. The sulfur-rich ejecta appear more asymmetrically distributed compared to oxygen or neon, a product of carbon-burning. In addition to the forbidden line emission from products of oxygen burning (S, Ar, Cl), we have also discovered H$\alpha$ and H$\beta$ emission from several knots of low surface brightness, fast-moving ejecta. 
The presence of fast-moving hydrogen points towards a progenitor that had not entirely shed its hydrogen envelope prior to the supernova. The explosion that gave rise to \1E0102 is therefore commensurate with a Type IIb supernova.
   \end{abstract}

   \keywords{Shock waves --- nuclear reactions, nucleosynthesis, abundances --- ISM: supernova remnants --- ISM: individual objects: SNR 1E0102.2-7219 --- Techniques: imaging spectroscopy}

\section{Introduction}
Core-collapse supernovae (CC SNe) mark the spectacular end of nuclear fusion processes within massive stars. Following the gravitational collapse of the stellar core and the subsequent explosion of the star as a supernova, freshly synthesized heavy elements are ejected at high velocity into the surrounding medium, thereby creating a supernova remnant (SNR). For the rare class of ``oxygen-rich'' (O-rich) supernova remnants (OSNRs), with typical ages ${\lesssim}3,000$\,years, the composition and spatial distribution of the stellar ejecta can be studied directly via the optical emission of its shock-heated atoms. These ejecta are detected at optical wavelengths primarily via the \oiii$\,\lambda\lambda$4959,5007 lines \citep[but see also][]{blair2000a}, following their encounter at a few 1000\,km\,s$^{-1}$ with the reverse shock \citep[][]{sutherland1995a}. OSNRs thus offer extraordinary windows into the explosion mechanisms and nucleosynthesis conditions of CC SNe, uniquely enabling the three-dimensional reconstruction of the ejecta morphology \citep[e.g.,][]{vogt2010a,vogt2011a,vogt2017a} and its composition, providing direct constraints for hydrodynamical explosion models \citep[e.g.,][]{wongwathanarat2017a}. Only a handful of such OSNRs are known: the most prominent examples include Cassiopeia A, Puppis A, and G292.2+1.8 in our own Galaxy, a remnant in the galaxy NGC 4449 (Blair et al. 1984), N132D and 0540--69.3 in the Large Magellanic Cloud (LMC), and 1E\,0102.2--7219 (hereafter 1E\,0102) in the Small Magellanic Cloud (SMC). The latter was discovered in  Einstein soft X-ray observations of the SMC by Seward \& Mitchell (1981), who found it to be the second brightest X-ray source in the SMC.  Dopita et al. (1981) classified 1E\,0102 as an OSNR. Here, we report the discovery of optical emission from high-velocity supernova ejecta in \1E0102 enriched in a number of elements not previously identified in \1E0102, namely hydrogen, sulfur, argon, and chlorine. 

\begin{figure*}[ht!]
\centerline{\includegraphics[scale=0.38, trim=10 70 120 60, clip]{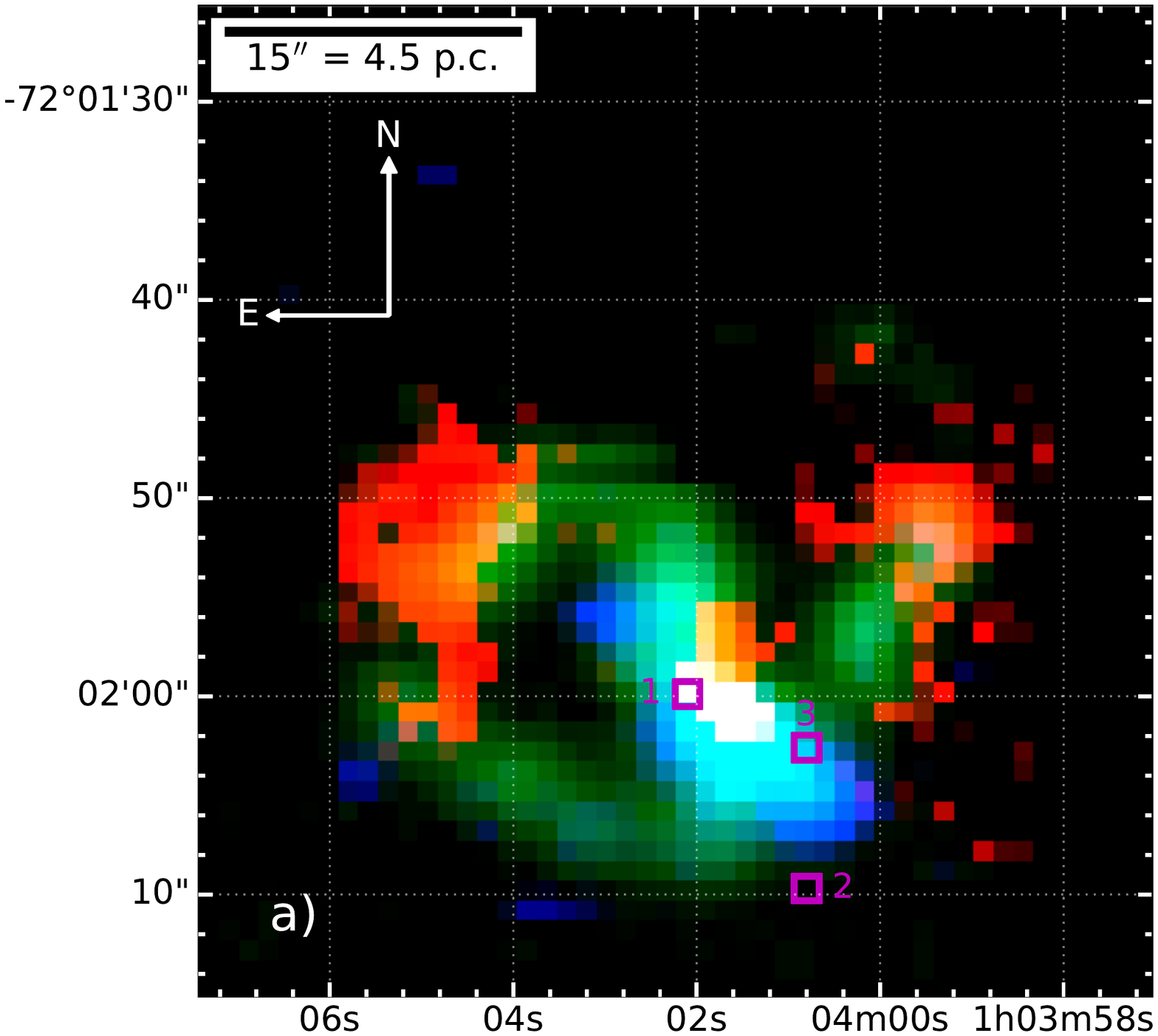}
\includegraphics[scale=0.38, trim=170 70 120 60, clip]{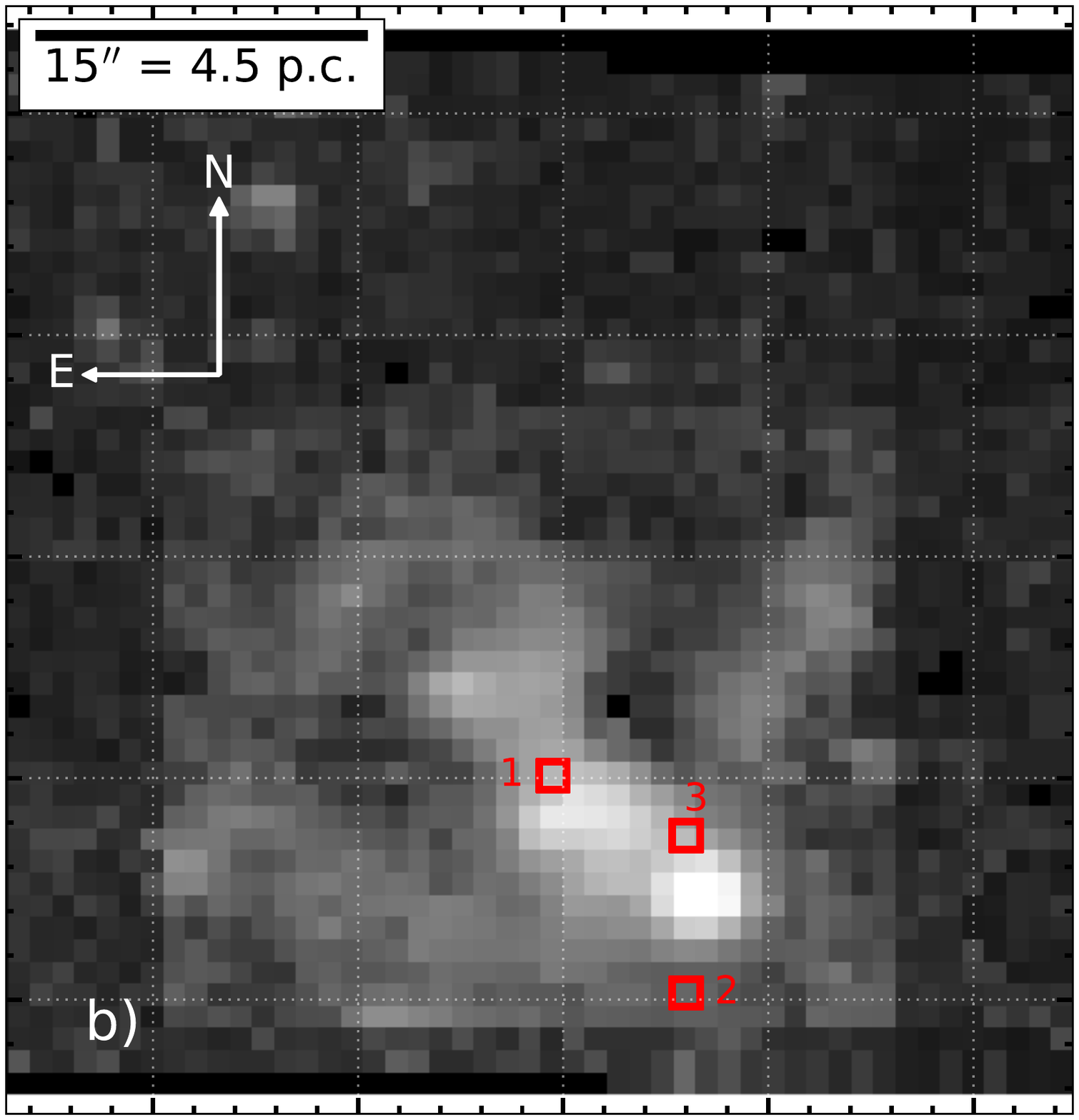}}

\centerline{\includegraphics[scale=0.38, trim=10 60 120 50, clip]{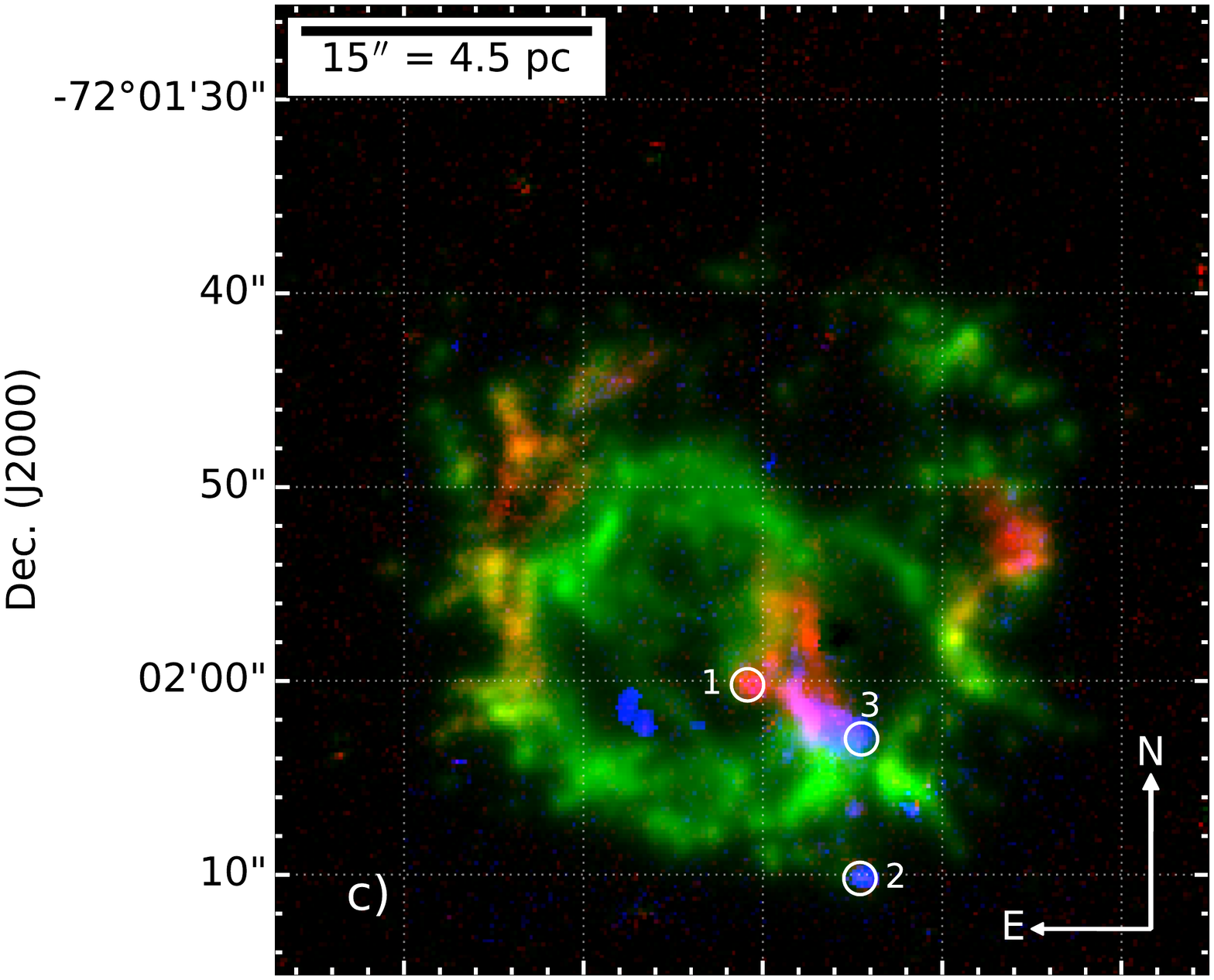}
\includegraphics[scale=0.38, trim=170 60 120 50, clip]{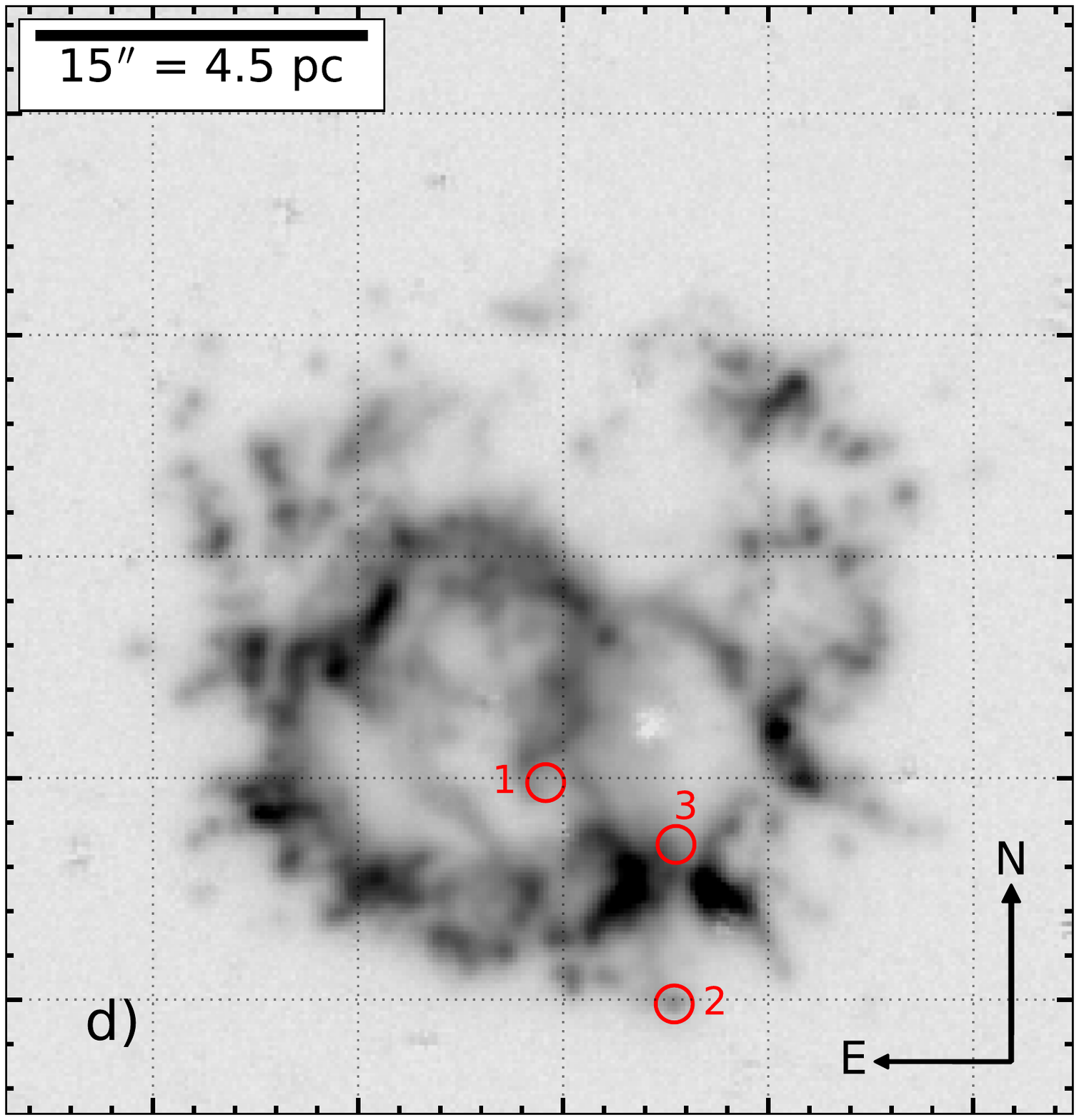}}

\centerline{\includegraphics[scale=0.38, trim=-10 5 130 50, clip]{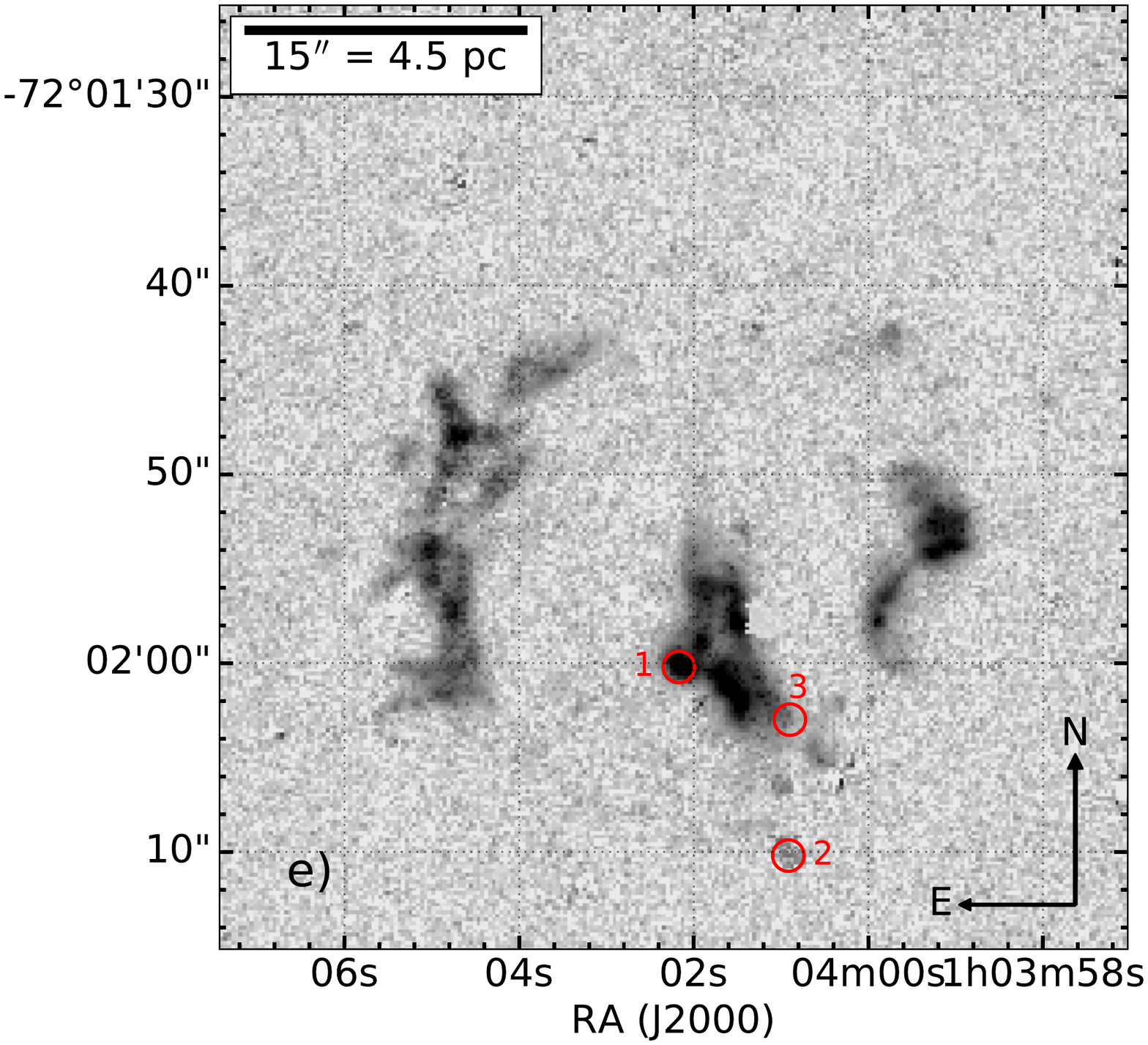}
\includegraphics[scale=0.38, trim=160 5 100 50, clip]{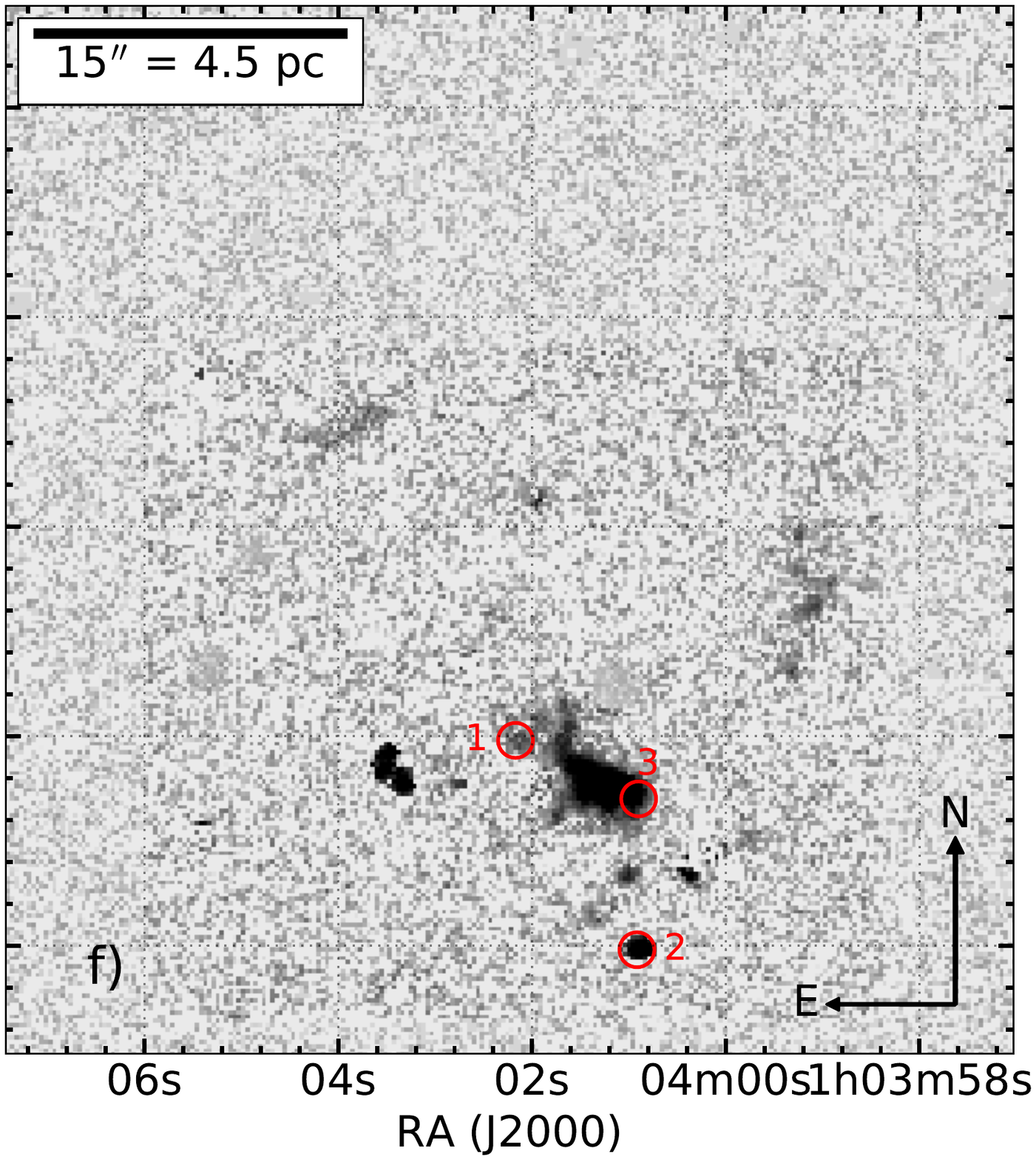}}

\caption{a) Pseudo RGB image of 1E0102 as seen by WiFeS, showing the emission from \sii\ (R), \oii\ (G) and \neiii\ (B). b) \neiii\ as seen by WiFeS. c) pseudo RGB image of 1E\,0102 as seen by MUSE, showing the emission from \sii\ (R), \oiii\ (G) and \ha\ (B). d) \oiii\ as seen by MUSE. e) \sii\ as seen by MUSE. f) \ha\  as seen by MUSE. All panels are based on projected line fluxes from Gaussian fits to the emission lines, which cover a range of Doppler shifts.  The small circles labeled 1, 2, and 3 mark extraction boxes for the spectra in Figures 2 and 3 and Table~1.}
\end{figure*}

\begin{figure*}[ht!]
\centerline{\includegraphics[scale=0.45, trim=20 10 140 10, clip]{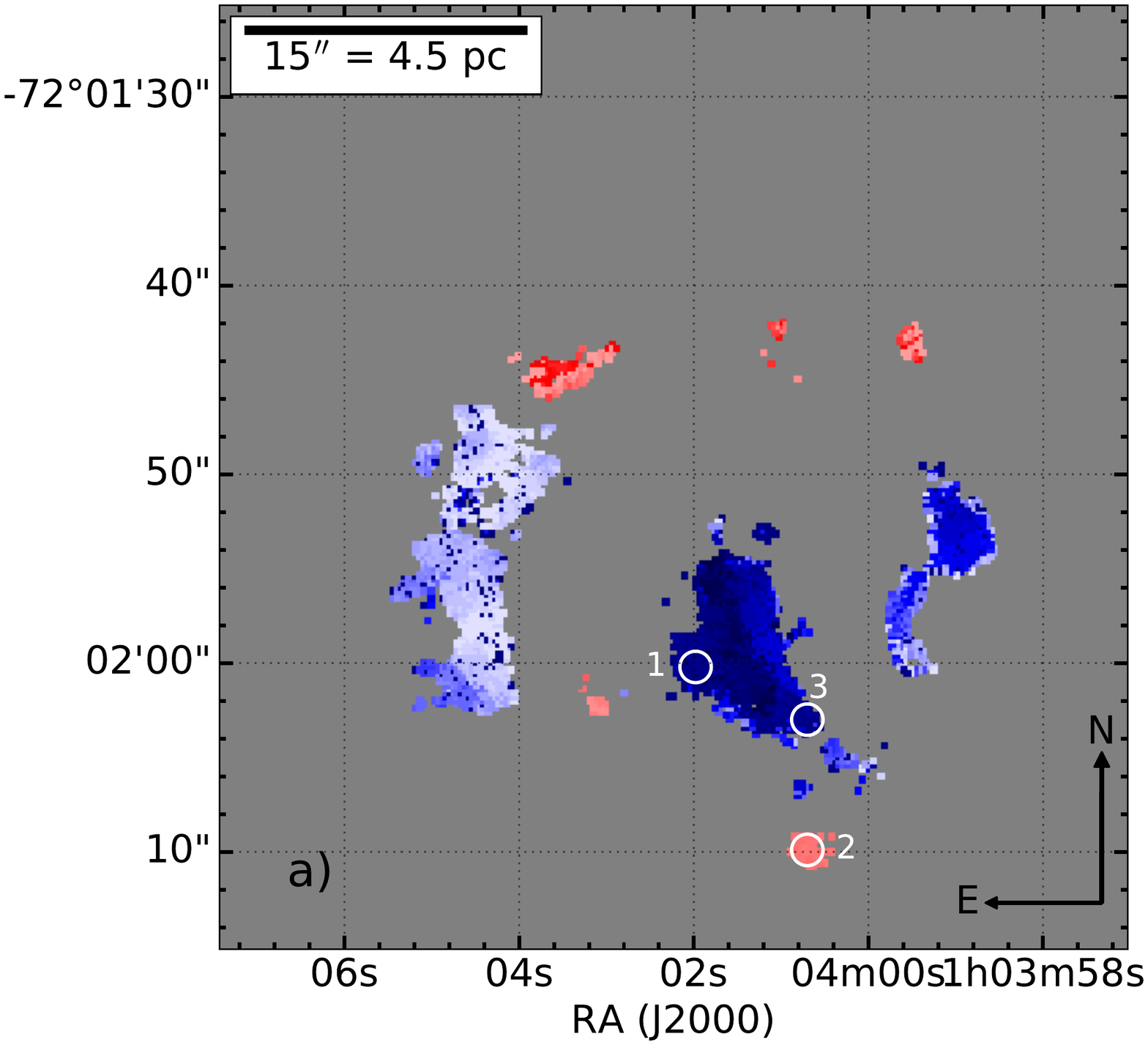}
\includegraphics[scale=0.45, trim=160 10 20 10, clip]{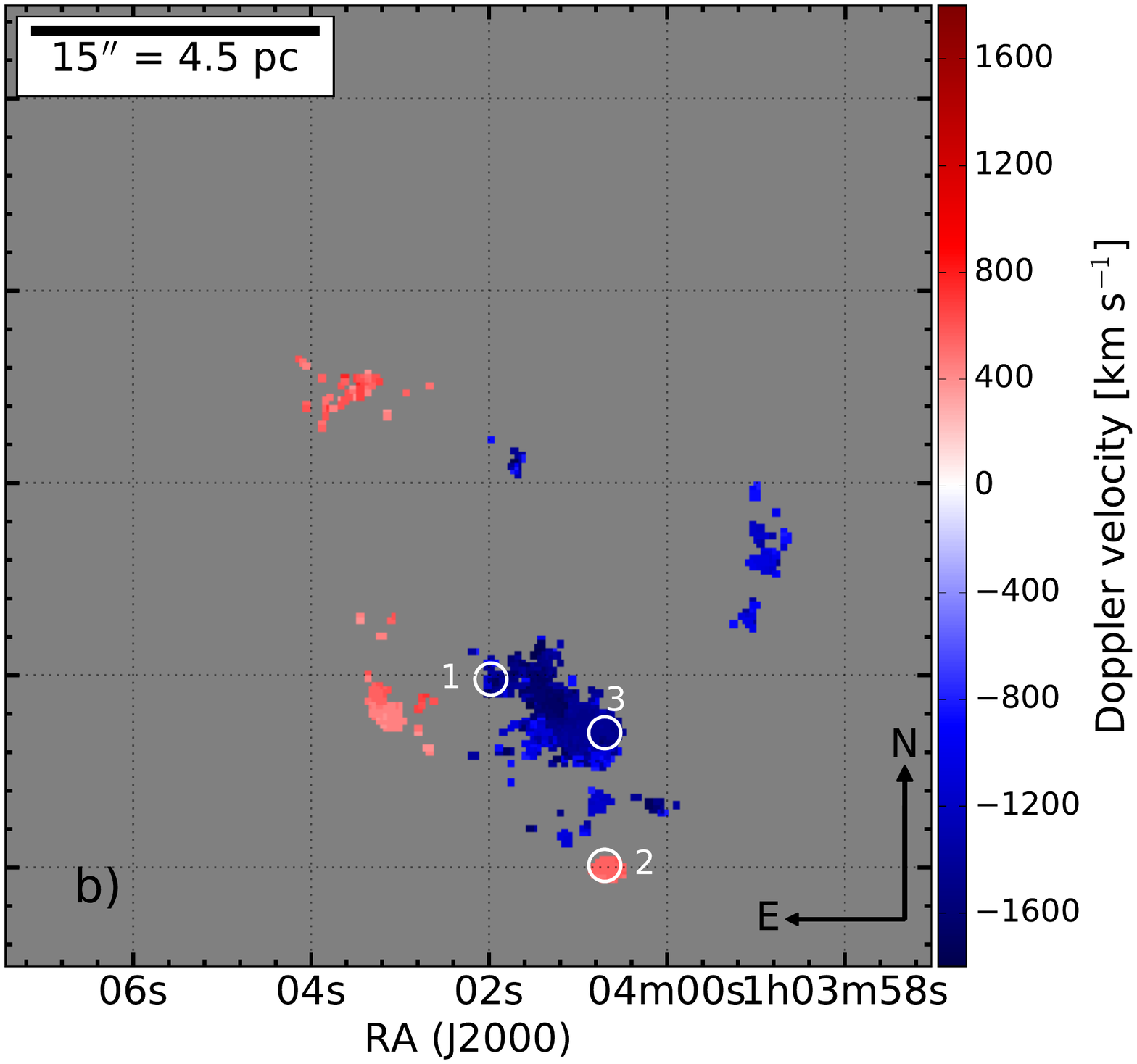}}
\caption{a) Extent of the \sii\ emission from Fig.~1e in Doppler velocity (R), b) Extent of the H$\alpha$ emission from Fig.~1f in Doppler velocity measured the local frame of the SNR, relative to the narrow emission lines.}
\end{figure*}


\begin{figure*}[ht!]
\centerline{\includegraphics[scale=0.4, trim=0 0 60 50, clip]{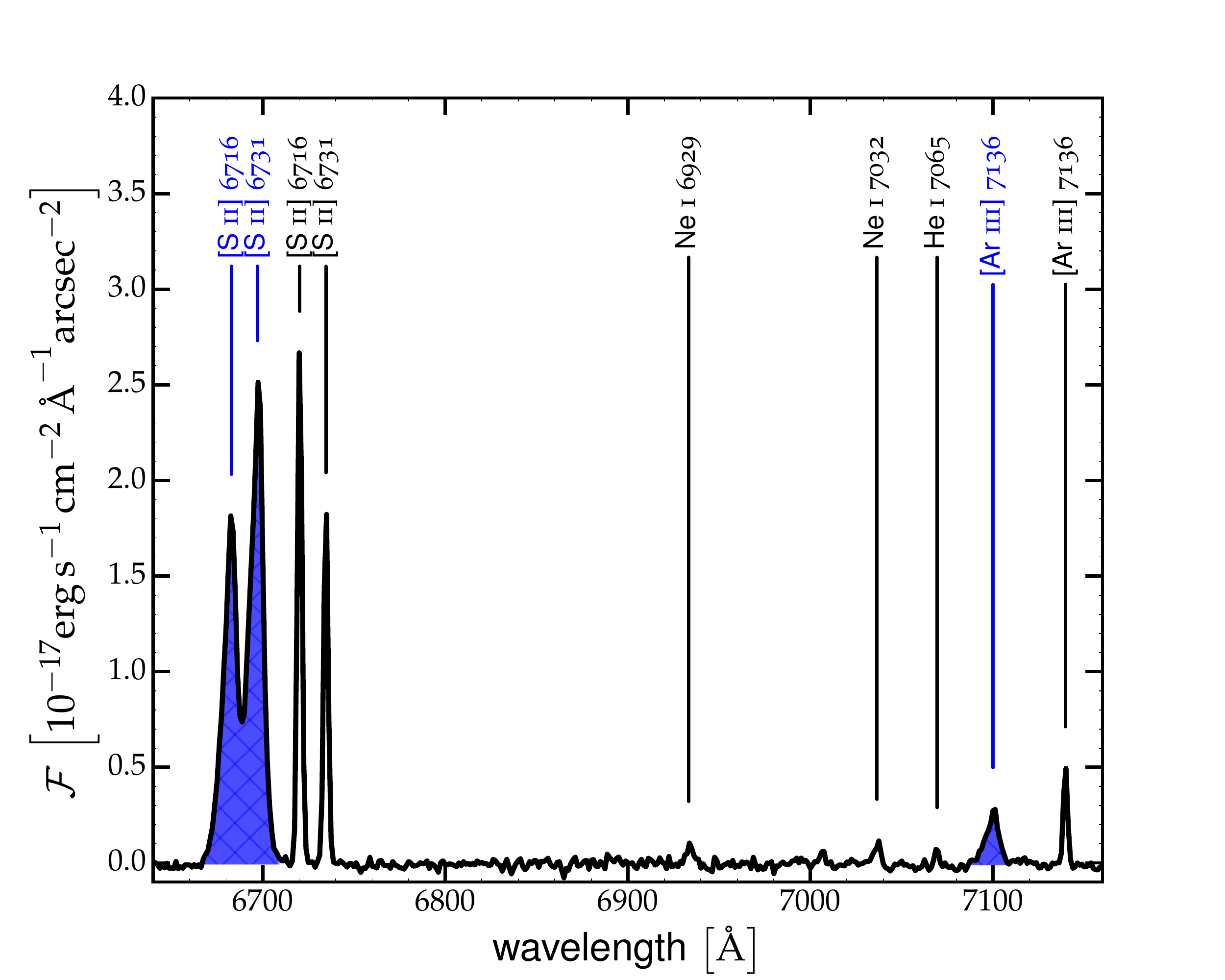}
\includegraphics[scale=0.4, trim=0 0 60 50, clip]{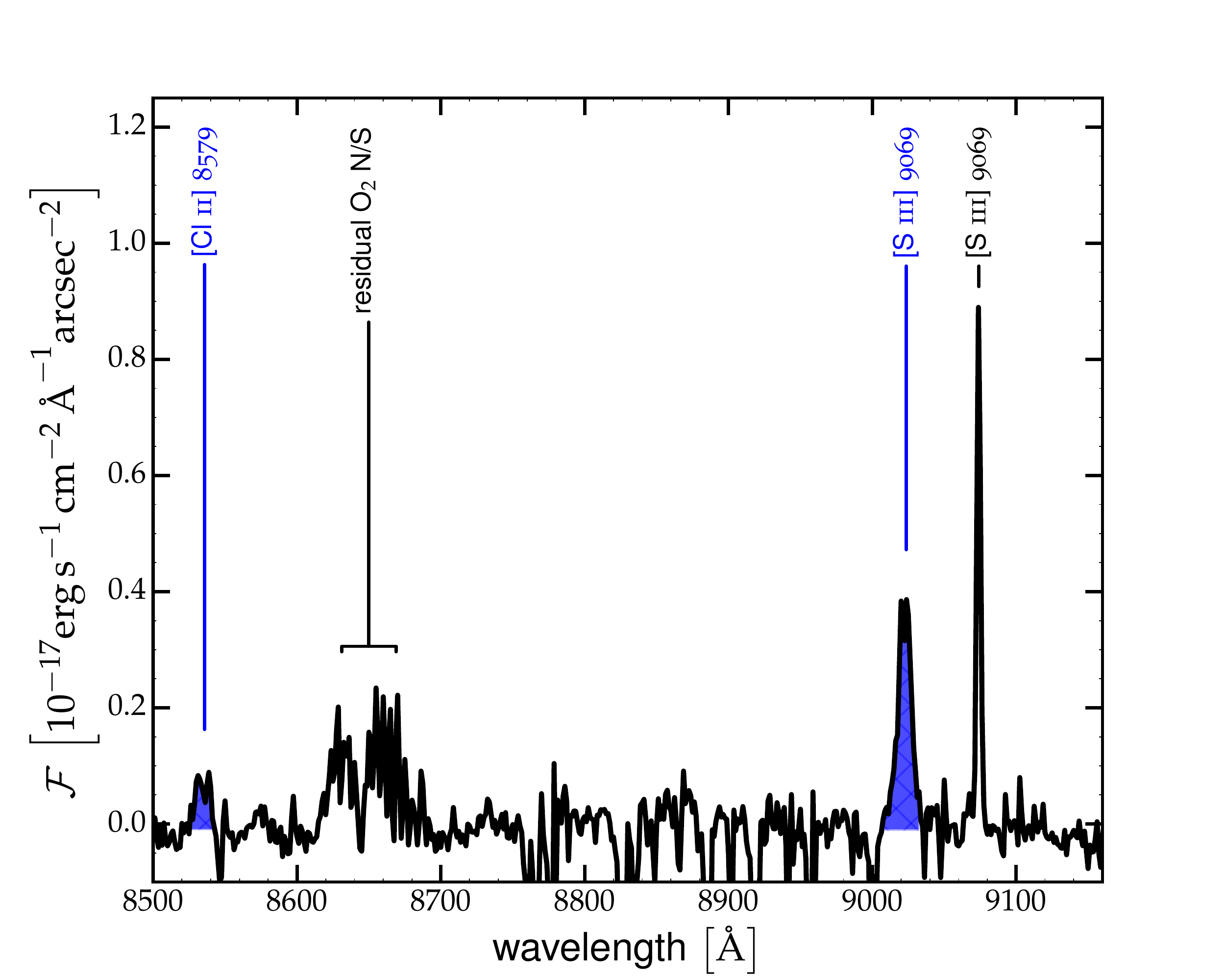}}
\caption{MUSE spectrum of region 1 (see Fig.~1) exhibiting strong, blue-shifted sulfur emission. The broad emission lines associated with the blue-shifted ejecta by $v=1655\,\mathrm{km}\,\mathrm{s^{-1}}$ (with respect to the narrow lines) are labeled in blue font, while the narrow emission lines at the Doppler velocity of the SNR are labeled in black font. }\label{fig:muse_eject_i}
\end{figure*}

\begin{figure*}[ht!]
\centerline{\includegraphics[scale=0.4, trim= 0 36 0 50, clip]{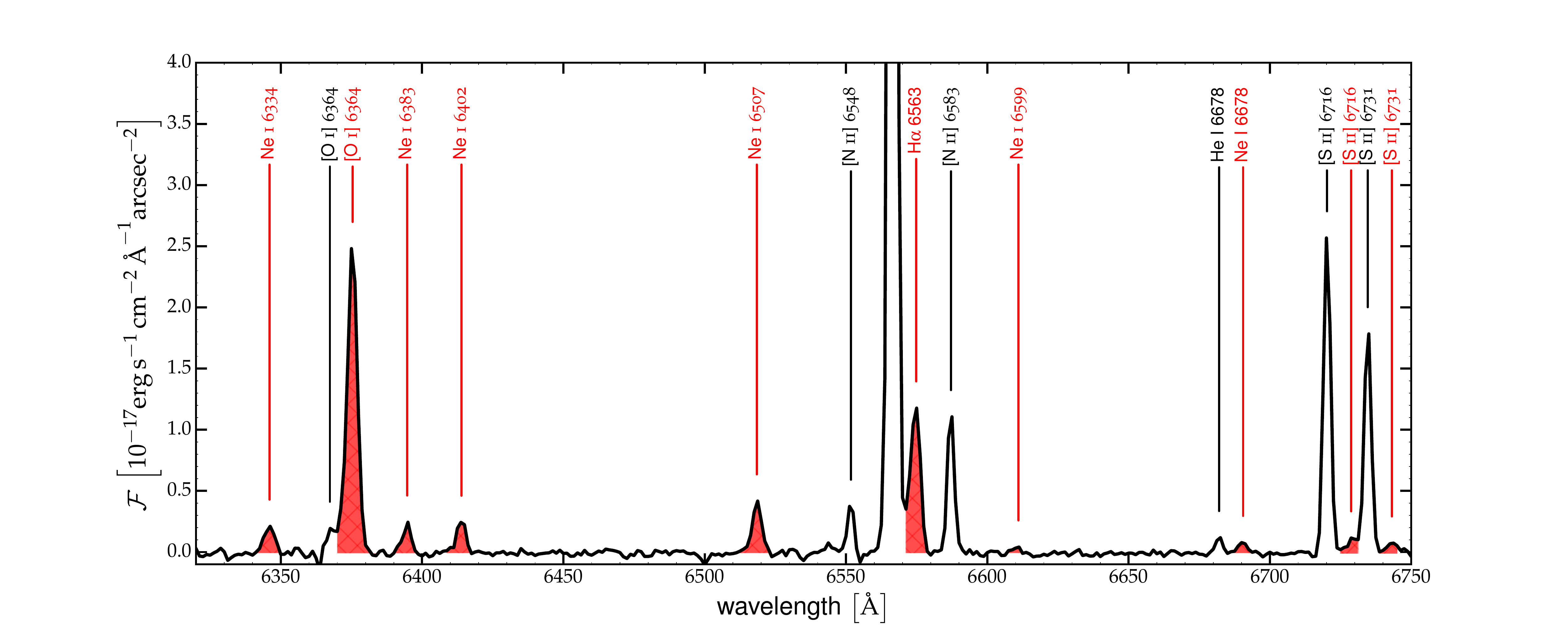}}
\centerline{\includegraphics[scale=0.4, trim= 0 6 0 50, clip]{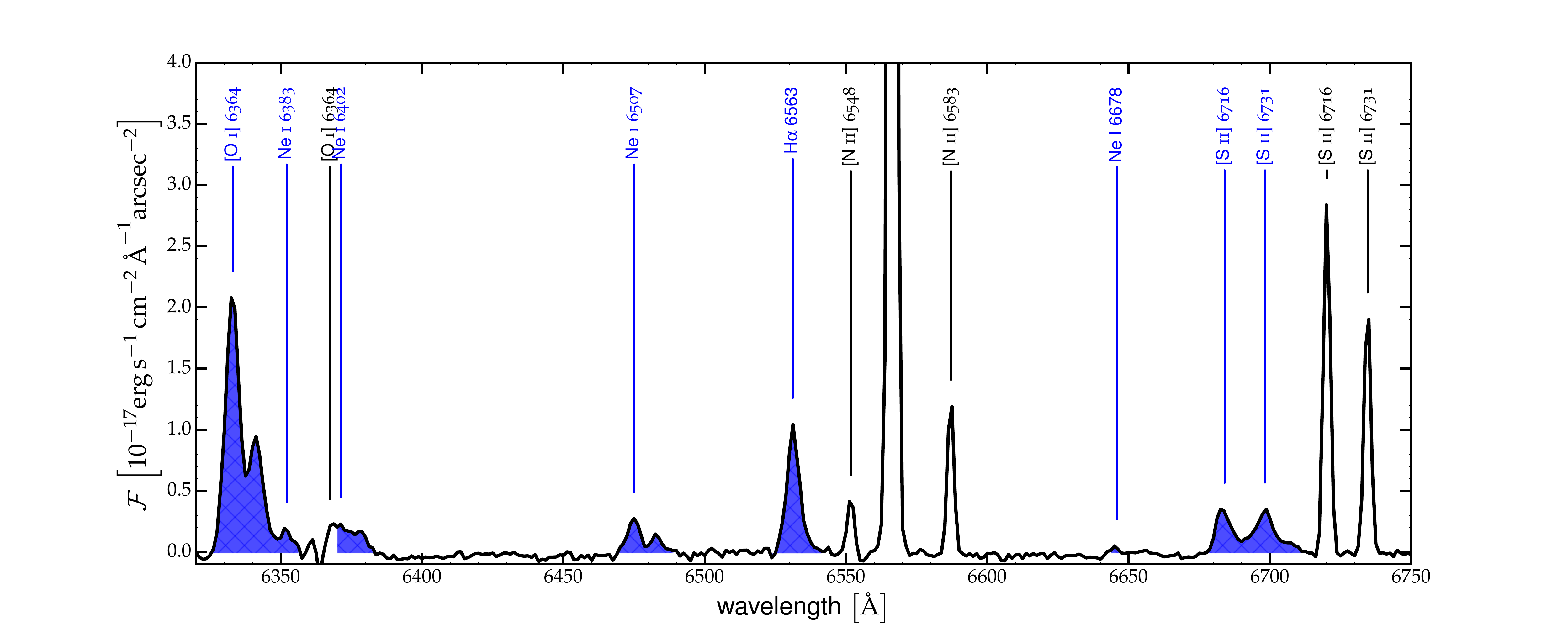}}
\caption{top: MUSE spectrum of H-rich knot in the south (region 2). The red lines are red-shifted by ${\sim}395\,\mathrm{km}\,\mathrm{s}^{-1}$ with respect to the narrow lines. bottom: MUSE spectrum of region 3. The stronger of the two high-velocity components in the bottom panel is blue-shifted by $1605\,\mathrm{km}\,\mathrm{s}^{-1}.$}\label{fig:muse_spec}
\end{figure*}

\begin{table*}[h]
\caption{Observed fluxes (with 1$\sigma$ statistical errors$^{a}$) and upper limits ($\mathcal{F}_\mathrm{obs}^i$) of selected emission lines in units of [$10^{-19}\, \mathrm{erg}\,\mathrm{s}^{-1}\, \mathrm{cm}^{-2}\,\mathrm{arcsec}^{-2}$], in regions 1 and 2. Observed ($\mathcal{I}_\mathrm{obs}^i$) and shock model line intensities ($\mathcal{I}_\mathrm{mod}^i$) are also given relative to H$\beta$=1.0. The observed fluxes quoted have been corrected with the \textsc{BRUTUS} code \citep{vogt2017c} for reddening with a \citet{fitzpatrick1999a} reddening law with $R_V=3.1$ and Galactic extinction \citep{schlafly2011a} with $A_B=0.134$ and $A_V=0.101$ extracted from the NASA/IPAC Extragalactic Database (NED).}
\vspace{1em}
\begin{centering}
\begin{tabular}{ l || c | c | c || c | c| c }
 \hline
line& $\mathcal{F}_\mathrm{obs}^1$ &$\mathcal{I}_\mathrm{obs}^1$
&$\mathcal{I}_\mathrm{mod}^1$
&$\mathcal{F}_\mathrm{obs}^2$
&$\mathcal{I}_\mathrm{obs}^2$
&$\mathcal{I}_\mathrm{mod}^2$\\ 
 \hline   
H$\beta$\,4861       & 9.0 \tiny{ $\pm$ 5.2}   & 1.0   & 1.0 & 29.0 \tiny{$\pm$ 2.5}     & 1.0   & 1.0 \\
\oiii\,4959          & 5956 \tiny{ $\pm$ 18}   & 662 \tiny{$\pm$ 382} & 564 & 4264.7 \tiny{$\pm$ 4.8} & 147 \tiny{$\pm$ 13} &  147 \\
\oiii\,5007          & 18037 \tiny{ $\pm$ 25}  & 2004 \tiny{$\pm$ 1158} & 1632& 12882.3  \tiny{ $\pm$ 7.5}& 444 \tiny{$\pm$ 38}  &  425 \\
\oi\,5577            & 17.5 \tiny{ $\pm$ 9.6}  & 1.94 \tiny{$\pm$ 1.55} & 0.13& 21.8  \tiny{ $\pm$ 1.3}   & 0.75 \tiny{$\pm$ 0.08}   &  0.08 \\
He\,\textsc{i}\,5876 & $<$2.7                  &$<$0.7& ...$^{c}$  & $<$4.5                    & $<$0.17        & ...$^{c}$\\
O\,\textsc{i}\,6157  & $<$2.8                  &$<$0.7& 1.48& 20.0  \tiny{$\pm$ 1.9}    & 0.69 \tiny{$\pm$ 0.09} & 1.34 \\
\oi\,6300            & 638.8 \tiny{ $\pm$ 6.4} & 71 \tiny{$\pm$ 41}   & 22.5& 951.4   \tiny{$\pm$ 2.0}  & 32.8 \tiny{$\pm$ 2.8}& 7.58\\
\oi\,6364            & 111.5 \tiny{ $\pm$ 9.2} & 12.8 \tiny{$\pm$ 7.2}  & 7.2 & 314.0    \tiny{$\pm$ 2.7} & 10.8 \tiny{$\pm$ 0.9}& 2.42\\
Ne\,\textsc{i}\,6507 & $<$7.3           &$<$1.9& ...$^{c}$ & 30.4 \tiny{ $\pm$ 5.7}    & 1.05 \tiny{$\pm$ 0.22} & ...$^{c}$ \\
\nii\,6548           & $<$1.3                  &$<$0.34& 0.01  & $<$3.6                    & $<$0.14          & 0.01 \\
H$\alpha$\,6563      & 35.4 \tiny{ $\pm$ 7.3}  & 3.93 \tiny{ $\pm$ 2.4}  & 4.1 & 140.1  \tiny{$\pm$ 2.2}   & 4.83 \tiny{$\pm$ 0.42} & 4.06 \\
\nii\,6583           & ...$^{b}$                      & ...$^{b}$    & 0.03& $<$1.9                    & $<$0.07 & 0.03 \\
\sii\,6717           & 353.4 \tiny{ $\pm$ 3.2} & 39.3 \tiny{ $\pm$ 23} & 39.2& 15.7 \tiny{$\pm$ 1.1}     & 0.54 \tiny{$\pm$ 0.06} & 0.45\\
\sii\,6731           & 558.0  \tiny{ $\pm$ 4.6}& 62.0 \tiny{ $\pm$ 35} & 31.5& 12.9  \tiny{$\pm$ 0.8}    & 0.44 \tiny{$\pm$ 0.05}& 0.32 \\
Ne\,\textsc{i}\,6929 & $<$6.7                  &$<$1.8& ...$^{b}$ & 11.6 \tiny{ $\pm$ 4.2}    & 0.40 \tiny{$\pm$ 0.15}& ...$^{c}$ \\
\ariii\,7136         & 62.2 \tiny{ $\pm$ 6.2}  & 6.9 \tiny{ $\pm$ 4.1}  & 6.9 & $<$4.5  & $<$0.17 & ...$^{c}$\\
\caii\,7291          & $<$7.7                  &$<$2.0                  & ...$^{c}$  & $<$9.0  & $<$0.34 & ...$^{c}$ \\          
\oii\,7319           & 1123 \tiny{ $\pm$ 11}   & 125 \tiny{ $\pm$ 72}   & 89.6& 975.0  \tiny{$\pm$1.9}    & 33.6 \tiny{$\pm$ 2.9} & 26.47\\       
\oii\,7330           & 571  \tiny{ $\pm$ 25}   & 63.4 \tiny{ $\pm$ 37} & 70.2& 735.0  \tiny{$\pm$ 2.2}   & 25.3 \tiny{$\pm$ 2.2} & 21.35 \\       
O\,\textsc{i}\,7774  & 122.3 \tiny{ $\pm$ 6.2} & 13.6 \tiny{ $\pm$ 7.9}  & 19.4& 301.4  \tiny{$\pm$ 1.5}   & 10.4 \tiny{$\pm$ 0.9} & 14.36 \\       
O\,\textsc{i}\,8446  & 17.6 \tiny{ $\pm$ 2.3}  & 2.0 \tiny{ $\pm$ 1.2}   & 11.9& 98.2 \tiny{$\pm$ 2.7}     & 3.39 \tiny{$\pm$ 0.31}& 2.71 \\
\clii\,8579          & 22.4 \tiny{ $\pm$ 4.9}  & 2.5 \tiny{ $\pm$ 1.5} & 2.48 & $<$7.7  & $<$0.29 & ...$^{c}$\\
\siii\,9069          & 121.6\tiny{ $\pm$ 4.6}  & 13.5 \tiny{ $\pm$ 7.8}  & 8.6  & $<$5.1& $<$0.19 & 0.13\\
O\,\textsc{i}\,9263  & 43.4 \tiny{ $\pm$ 3.8}  & 4.8 \tiny{ $\pm$ 2.8}  & 9.0  & 169.5 \tiny{$\pm$ 1.8}    & 5.84 \tiny{$\pm$ 0.51} & 2.65\\
 \hline
\end{tabular}\\
\end{centering}
\vspace{1em}
\textbf{Notes:}\\
  $^{a}${We are quoting 1$\sigma$ statistical errors of the Gaussian fit. Note that in cases of strong lines with very high signal to noise, such as e.g. \oiii, systematic uncertainties dominate and the statistical errors under-estimate the total uncertainty of the observed line flux.}\\
$^{b}${A meaningful upper limit could not be determined in this case as the Doppler shift puts \nii\,6583 on top of H$\alpha$. }\\
  $^{c}${Not included in the \textsc{MAPPINGS V} shock model calculations, either because the upper limit does not lead to meaningful constraints or the line is not included in the code.}
\end{table*}


\section{Detection of sulfur-rich ejecta}
We observed \1E0102 with the WiFeS Integral Field Spectrograph on the 2.3\,m telescope of Siding Spring Observatory on 2016 August~7 and 2016 August~8.  In the reduced datacube we unambiguously detected faint, highly Doppler-shifted \sii $\lambda\lambda6716,6731$ emission (see Fig.~1a), associated with the ejecta of this OSNR \citep[see also][]{seitenzahl2017a}. 
Following this discovery, we performed follow-up observations of  \1E0102 using the Multi-Unit Spectroscopic Explorer (MUSE) in Service Mode at the Very Large Telescope on the night of 2016 October 7, as part of Director Discretionary Time (DDT) program 297.D-5058 (P.I.: F.P.A. Vogt) for $9\times900$\,s exposures on-source. For details of the observations and the data reduction see \citet{vogt2017b}.
In the \sii\ images of \1E0102 extracted from the MUSE datacube, the smaller pixel scale of the MUSE data compared to WiFeS (0.2$^{\arcsec}$ per spaxel vs 1$^{\arcsec}$ per spaxel) allows a much finer discernment of the spatial distribution of S-rich ejecta relative to the O-rich ejecta (e.g., Fig.~1a,c). The MUSE imagery shows that the \sii\ emission is patchy and asymmetric, aligning with localized segments of \oiii-emitting material along the eastern, western and central regions. The sulfur-rich ejecta is detected at Doppler shifts ranging from $-$1776\,\kms\ to $+$958\,\kms\ (Fig.~2a). In some of the brighter regions,  \siii\ $\lambda$9069, \ariii $\lambda$7136, and \clii\ $\lambda$8579 are also detected at consistent Doppler velocities (see Fig.~3).  

These data are the first detection in the optical of the long sought-after \citep[][]{lasker1991a,blair2000a} products of oxygen-burning in the ejecta of the supernova that gave rise to 1E\,0102. Fitted line fluxes (and some upper limits) of the brightest sulfur-rich knot (region 1 in Fig.~1) are given in Table~1 for a selection of emission lines.

\section{Discovery of the fast-moving hydrogen}
Aside from the sulfur-rich emission, we also discovered several distinct high velocity knots emitting in the Balmer lines of hydrogen (see Fig.~1f, Fig.~2b, and Fig 4.). This Balmer emission is also observed from both blue-shifted and red-shifted material.  The velocity shifts of the H$\alpha$ and H$\beta$ lines agree with the Doppler shifts of other prominent identified lines from the same extraction region, giving confidence in the line identification and that the knots are indeed high-velocity hydrogen-rich ejecta.  The hydrogen-rich ejecta is detected at Doppler shifts ranging from $-$1785\,\kms\ to $+$786\,\kms (Fig.~2b).  Interestingly, the distribution of the Balmer-line emission somewhat correlates with the \sii\ and to a lesser extent \oiii\ emission (cf. Fig.~1d,e,f and Fig.~2a,b). Comparison with the WiFeS data shows that emission from these lines also coincides with \neiii $\lambda$3889 near the center (see Fig.~1b). Fitted line fluxes (and some upper limits) of the prominent Balmer-line emitting knot in the south (region 2 in Fig.~1) are given in Table~1 for a selection of emission lines. 

\section{Shock model calculations}
\label{sec:models}
We have constructed cloud shock models for region 1 and 2 using the \textsc{MAPPINGS V} code with self-consistent pre-ionisation as described by \citet{sutherland2017a}. The structure of these O-rich shocks is quite different from those of normal plasmas \citep{itoh81a,dopita84a,sutherland95b}. First, in the post-shock region, the plasma cools very quickly, so that colliesional ionisation is never reached, and the plasma cools below 1000\,K before significant recombination to un-ionised states has occurred. Second, the ionising photon production in these shocks is high, so that they produce photoionised precursors. In these the time-dependence of the ionisation is very important. Initially, the incoming plasma is exposed to hard photons left over from maintaining ionisation in the precursor, and the fractional ionisation is very low. As a consequence, the photo-heating much exceeds to cooling provided by the few electrons, and the plasma is strongly super-heated. This is the region which dominates the emission of the low-ionisation species. As the plasma becomes more fully ionised in its passage towards the shock front, the cooling dominates the photo-heating, and the plasma falls towards its equilibrium temperature of $\sim 100-200$\,K. In these shocks, the forbidden line emission from the leading edge of the photoionised precursor zone may be as important, or more important than the forbidden line emission from the cooling region of the shock itself.

To model regions 1 and 2 we have used the iterative approach described by  \citet{sutherland2017a} to model both regions. The observed ratios of \oi\ to \oii\ to \oiii\ constrain the shock velocities to be relatively low, 30 -- 100 \kms, consistent with the model in which the O-rich knots arise in relatively dense ejecta passing through the reverse shock. 
The best fit model requires a mixture of shock velocities to be present, as might be expected if there exist variations in the pre-shock density in the shocked cloud. For Region 1, we mixed shock velocities between 50 and 90 \kms\ to produce a satisfactory fit, while for Region 2 the range of velocities required to produce a good fit was smaller; 40 -- 50 \kms. For the Region 1 best-fit model, the pre-shock density is ${\sim}1.8\times10^{-21} \mathrm{g}\, \mathrm{cm}^{-3}$, while for Region 2, the pre-shock density is ${\sim}5.4\times10^{-22} \mathrm{g}\, \mathrm{cm}^{-3}$. Those correspond to ionic densities of $230\,\mathrm{cm}^{-3}$ and $22.2\,\mathrm{cm}^{-3}$, respectively.

In all of the models the magnetic parameter $\alpha$, the ratio of the magnetic pressure to the gas pressure in the precursor was taken to be unity. We note that the assumption of magnetic pressure equal to the gas pressure in the shock precursor is one of energy equipartition, which will be strictly valid in a turbulent precursor medium. However, since the gas in the far precursor region is at very low temperature, a few tens of degrees K, this assumption is almost equivalent to the low-magnetic field limit in the shock. The precursor magnetic field would have to be much higher to appreciably affect the results presented here, which seems to be unphysical. Therefore, our results are only weakly dependent on the exact value of the magnetic field.

Since the spectrum is dominated by emission from oxygen, we adjusted the abundances of the other observed elements relative to O, For Region 1 we estimate the following abundances 
$\log(\mathrm {H/O}) \sim -1.4$, 
$\log(\mathrm {N/O}) \lesssim -4.5$,
$\log(\mathrm {S/O}) = -1.8$,
$\log(\mathrm {Cl/ O}) = -3.5$, and 
$\log(\mathrm {Ar/O}) = -2.3$.
While for  Region 2, we derive 
$\log(\mathrm {H/O}) = -1.0$,
$\log(\mathrm {N/O}) \lesssim  -5.0$, and 
$\log(\mathrm {S/O}) = -4.5$.
Typical errors on these figures are of order 0.2 -- 0.3 dex.

\section{Discussion and Conclusions}
\citet{chevalier2005a} argue that \1E0102 is the remnant of a Type IIL/b supernova with a progenitor that lost most (but not all) of the H envelope before the explosion. The very low surface brightness of the Balmer emission makes a Type IIL progenitor implausible, but a Type IIb supernova type for 1E\,0102 would place this remnant in the same category as the (in many regards similar) OSNRs Cas~A and Puppis~A.  Indeed, fast hydrogen-rich knots, similar to those we report here, were discovered both in Cas~A -- known to stem from a Type IIb SN from light echo studies \citep{krause2008a} -- by \citet{fesen1988a} and \citet{fesen1991a}, and Puppis~A \citep{winkler1985a}, which \citet{chevalier2005a} also assigned to a Type IIL/b SN progenitor. The high velocities of these features in Cas A ($\sim$6000\,\kms) led \citet{fesen1988a} and \citet{fesen1991a} to the conclusion that the hydrogen originated in the outermost photospheric region of Cas A's progenitor, where the outflow velocities are highest.  We note that the hydrogen-rich features in 1E\,0102 exhibit somewhat lower Doppler shifts, up to about -1785\,\kms. 

Notably, we do not observe any emission from \nii\ (see upper limits for region 2 in Table~1), neither associated with the Balmer line emitting knots as is the case for Cas A \citep{fesen1988a} nor in N-rich filaments as observed for Puppis A \citep{winkler1989a}. For Cas A, \nii$\lambda6583$/H$\alpha > 6$ in many of the FMF, which in turn implies a nitrogen abundance of at least 10 times solar \citep{fesen1988a,fesen2001a}. In stark contrast, in region 2 we find \nii$\lambda6583$/H$\alpha < 0.014$ (see Table~1), which is at least a factor of $6/0.014=429$ lower than for the FMF in Cas~A. This does not, however, imply a low nitrogen abundance relative to hydrogen. Our shock models indicate that very low shock speeds ($40-50\,\mathrm{km}\,\mathrm{s}^{-1}$) are required to reproduce the observed spectrum in region 2. The approximately solar nitrogen to hydrogen ratio assumed for this calculation results in a line intensity of \nii$\lambda6583$ well below the observational upper limit and even a super-solar nitrogen to hydrogen ratio of twice the solar value, ${\sim}8\times10^{-5}$ \citep[e.g.][]{lodders2003a} could be accommodated.   

In addition to the new detections of fast-moving Balmer emission and S-rich material, we have also detected emission from argon (\ariii $\lambda$7136) and chlorine (\clii $\lambda$8579). 
The abundances for our shock model that reproduces the spectrum for region 1 reasonably well (see Table~1 and Section~\ref{sec:models}) indicate a strong enhancement relative to H of O, S, Ar, and Cl.
Using \citet{russell1992a}, we find enrichments in region 1 of $10^{5.37}{\approx}235,000$ times the ISM abundance of the SMC for oxygen, $10^{5.01}{\approx}100,000$ times SMC for sulfur, $10^{5.20}{\approx}160,000$ times the ISM abundance of the SMC for chlorine, and $10^{5.29}{\approx}195,000$ times SMC for argon.
The enhancement of products of oxygen-burning (e.g., S, Cl, and Ar) indicates that nuclear fusion has partially proceeded past the previously known elements associated with explosive burning of the neon-carbon shell, i.e., C, Ne, O, and Mg. The enhancement in region 2 is less pronounced, with an over-abundance of sulfur relative to hydrogen of ${\sim}200$ times the SMC value \citep{russell1992a}, and argon and chlorine not detected. 

The sulfur-rich ejecta appears more asymmetrically distributed than oxygen/neon and correlates with the hydrogen emission. 
The discovery of the fast  hydrogen containing knots implies that the progenitor star retained part of its hydrogen envelope up moments before the supernova explosion. 
This makes a Wolf-Rayet progenitor of a Type Ib or Type Ic supernova unlikely for 1E\,0102. 
Based on the newly discovered fast-moving Balmer-line emission, we therefore favor a Type IIb supernova progenitor that was stripped of most -- but not all -- of its hydrogen envelope, possibly via interaction with a close companion star. 

\begin{acknowledgements}
This research has made use of the following \textsc{python} packages: \textsc{statsmodel} \citep{seabold2010a}, 
\textsc{matplotlib} \citep{hunter2007a}, \textsc{astropy}, a community-developed core Python package for Astronomy \citep{AstropyCollaboration2013}, \textsc{aplpy}, an open-source plotting package for \textsc{python} hosted at \url{http://aplpy.github.com}, \textsc{mpfit} \citep[][]{markwardt2009a, more1978a} and \textsc{mayavi} \citep[][]{ramachandran2011a}.
This research has also made use of \textsc{montage}, funded by the National Science Foundation under Grant Number ACI-1440620 and previously funded by the National Aeronautics and Space Administration's Earth Science Technology Office, Computation Technologies Project, under Cooperative Agreement Number NCC5-626 between NASA and the California Institute of Technology, of the \textsc{aladin} interactive sky atlas \citep{bonnarel2000a}, of SAOImage DS9 \citep{joye2003a} developed by Smithsonian Astrophysical Observatory, and of NASA's Astrophysics Data System. This research has made use of the NASA/IPAC Extragalactic Database (NED) which is operated by the Jet Propulsion Laboratory, California Institute of Technology, under contract with the National Aeronautics and Space Administration. Based on observations made with ESO Telescopes at the La Silla Paranal Observatory under programme ID 297.D-5058[A]. IRS acknowledges support from the Australian Research Council Grant FT160100028. AJR acknowledges support from the Australian Research Council through project numbers CE110001020 (CAASTRO) and FT170100243.
\end{acknowledgements}


\end{document}